\documentclass[a4paper,11pt]{article}
\usepackage{pos}
\usepackage{wrapfig}

\title{Targeting 100-PeV tau neutrino detection with an array of phased and high-gain reconstruction antennas}
\ShortTitle{GRAND-BEACON}

\author*[a]{Stephanie Wissel}
\author[a]{Andrew Zeolla}
\onbehalf{for the BEACON Collaboration}
\author[b,c]{Tim Huege}
\author[a,d,e]{Kumiko Kotera}
\author[d,f]{Olivier Martineau}
\onbehalf{for the BEACON and GRAND Collaborations \\{\normalsize \normalfont(a complete list of authors can be found at the end of the proceedings)}\\}

\affiliation[a]{Depts of Physics \& Astronomy and Astrophysics, CMA, IGC, Penn State, University Park, PA, USA}
\affiliation[b]{Inst. for Astroparticle Physics (IAP), Karlsruhe Inst. of Technology, 76021 Karlsruhe, Germany}
\affiliation[c]{Astrophysical Inst., Vrije Universiteit Brussel, 1050 Brussels, Belgium}
\affiliation[d]{Sorbonne Université, CNRS, UMR 7095, Inst. d'Astrophys. de Paris (IAP), Paris, France}
\affiliation[e]{Phys. Dept., Vrije Universiteit Brussel, Pleinlaan 2, 1050 Brussels, Belgium}
\affiliation[f]{Laboratoire de Physique Nucléaire et des Hautes Energies (LPNHE), Inst. d'Astrophys. de Paris (IAP), Paris, France}
\emailAdd{wissel@psu.edu}

\abstract{Neutrinos at ultrahigh energies can originate both from interactions of cosmic rays at their acceleration sites and through cosmic-ray interactions as they propagate through the universe. These neutrinos are expected to have a low flux which drives the need for instruments with large effective areas. Radio observations of the inclined air showers induced by tau neutrino interactions in rock can achieve this, because radio waves can propagate essentially unattenuated through the hundreds of kilometers of atmosphere. Proposed arrays for radio detection of tau neutrinos focus on either arrays of inexpensive receivers distributed over a large area, the GRAND concept, or compact phased arrays on elevated mountains, the BEACON concept, to build up a large detector area with a low trigger threshold. We present a concept that combines the advantages of these two approaches with a trigger driven by phased arrays at a moderate altitude (1 km) and sparse, high-gain outrigger receivers for reconstruction and background rejection. We show that this design has enhanced sensitivity at 100 PeV over the two prior designs with fewer required antennas and discuss the need for optimized antenna designs.}

\FullConference{10th International Workshop on Acoustic and Radio EeV Neutrino Detection Activities (ARENA2024)\\
11-14 June 2024\\
The Kavli Institute for Cosmological Physics, Chicago, IL, USA\\}

\begin{document}
\maketitle

\section{Introduction}

Astrophysical neutrinos are expected to be produced at energies extending beyond the PeV scale, at ultra-high energies (UHE)~\cite{Guepin:2022qpl}. Neutrinos at these energies are excellent messenger particles which point back to their sources and indicate the presence of non-thermal cosmic ray acceleration, both nearby and in the distant past. 

A promising technique for discovering UHE neutrinos looks for air showers resulting from neutrinos as they skim the edge of the Earth's limb. Tau neutrinos can interact in the Earth to produce tau leptons which can decay in the atmosphere producing an extensive air shower. These tau lepton-induced air showers are concentrated towards the horizon because the probability that a tau lepton will exit with sufficient energy increases towards the horizon. This process is dominated by tau neutrinos because they can regenerate -- meaning that tau neutrinos undergoing charged current interactions produce tau leptons and vice versa with minimal energy loss as they propagate through the Earth, while electrons would be absorbed and the muon decay length is long enough that they only become relevant at satellite altitudes.
In recent years, several concepts targeting Earth-skimming tau neutrinos using radio detection techniques have emerged.
The radio technique is uniquely suited for detecting these Earth-skimming air showers, particularly because radio emission from the distant, inclined showers can propagate from the horizon without significant attenuation. 

The Giant Radio Array for Neutrino Detection (GRAND) concept uses a large sparse radio array of autonomously triggered antennas to target the large radio footprints of inclined showers~\cite{GRAND}. A total of 200,000 antennas in 10 arrays of 20,000 are planned to build nearly full sky coverage. The advantage of such a design is that the sparse layout covers a large area on a the ground making reconstruction, background rejection, and neutrino identification readily feasible. Advanced triggers can take advantage of the differences in the neutrino and cosmic ray radio characteristics~\cite{Correa:2023maq, Kohler:ARENA2024}. 

The Beamforming Array for COsmic Neutrinos (BEACON) concept uses compact arrays of phased antennas placed on a high elevation mountain~\cite{Wissel_2020}. Because the detectors are placed at a high prominence ($>2$\,km) above the horizon, the design maximizes the visible geometric array available to a ground experiment. Multiple beams formed by delaying-and-summing the signal from the antennas into phased arrays can lower the trigger threshold of the array and tune the triggering region. In the beam pointed in the direction of the signal, the signal adds linearly with the number of antennas, $N_{\rm ant}$, included in the beam while the incoherent noise increases as $\sqrt{N_{\rm ant}}$. These effects combined require fewer total channels such that the full BEACON design includes 1000 stations of 10 phased antennas each. 

\section{Concept for the 100 PeV Tau Neutrino Detector}

In this proceeding, we aim to further optimize a radio-based Earth-skimming tau neutrino detector, building on the strengths of both GRAND and BEACON. Our goal is to develop an experiment optimized for the 100\,PeV energy scale where the flux is expected to be higher even if the exit probability is lower. Design considerations include the detector elevation above the nearby horizon, or the prominence; the number of antennas included in a phased array that provides a trigger; the antenna frequency range, gain, and beam pattern which can vary for antennas included in the trigger and antennas used for reconstruction; the fraction of the sky noise that is visible to the antennas; and how much of the sky noise is reflected by the ground.

The sensitivity of a radio tau neutrino detector is driven by the total geometric area visible to the triggering portion of the array and the signal-to-noise of the RF emission from the events. The SNR can be enhanced in two ways: by increasing the signal relative to the noise or by decreasing the noise temperature. The former is accomplished here via phasing, while the later can be accomplished if the sky fraction, $r$, or Fresnel reflection coefficient, $\mathcal{F}$, is adjusted, because the antenna temperature, $T_{\rm ant}$ depends on:
\begin{equation}\label{eqn:ant_temp}
    T_{\rm ant} = (r + (1-r)\mathcal{F})T_{\rm gal} + (1-r)T_{\rm ground} + T_{\rm sys}
\end{equation}
where the irreducible noise is due to the galactic noise $T_{\rm gal}$, the blackbody radiation of the ground $T_{\rm ground}$, and the system temperature from the active components in the signal chain $T_{\rm sys}$. 

The frequency range and antenna gain are important but nuanced factors in the optimization procedure. Galactic noise, $T_{gal}$, rapidly decreases a power law with frequency and becomes comparable to the ground temperature in the 100-200 MHz range. At frequencies below roughly 100\,MHz, the radio emission follows a filled-in Cherenkov cone, making an instrument more sensitive to a broader range of incoming neutrino geometries. At higher frequencies, the signal is more strongly peaked at the Cherenkov angle, which can be beneficial for enhancing the sensitivity at low energies~\cite{Wissel_2020, GRAND}. 

\begin{figure}[b]
    \centering
    \includegraphics[width=0.7\linewidth]{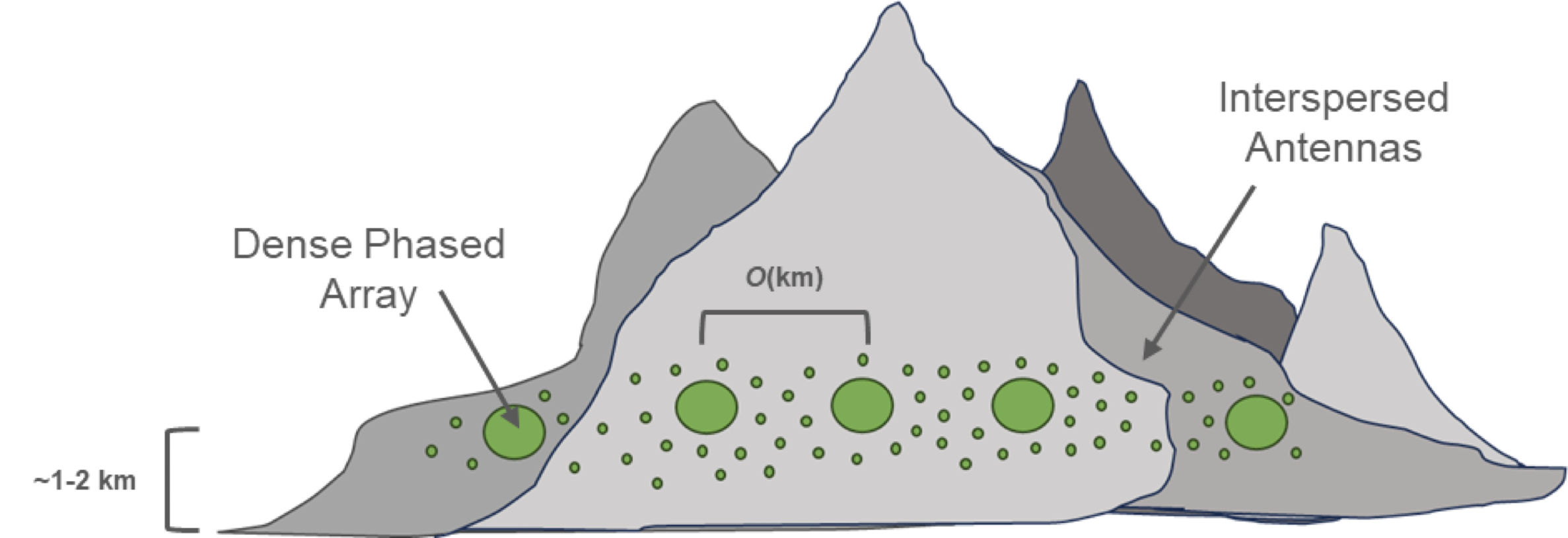}
    \caption{Conceptual diagram for a hybrid array formed with densely packed phased array antennas for lower thresholds and triggering and a more sparse array for reconstruction and background rejection.}
    \label{fig:conceptdiagram}
\end{figure}

The frequency range and gain chosen are also driven by practical concerns: lower frequency antennas are typically larger and/or electrically short, while higher frequency designs are smaller and can more readily be designed with higher antenna gain.  Enhancing the gain of the antenna narrows the beam and amplifies the signal if the antenna is pointed at it. Given that the showers are narrowly beamed towards the horizon, this can be an advantage. The noise is integrated over the full antenna beam and so the sky fraction $r$ is the relevant parameter rather than the gain. But by pointing the beam lower into the horizon where the signal is expected to peak and sky fraction is reduced, the SNR can be increased. For the fiducial design presented here, we chose the 30-80\,MHz range, but emphasize that this could be further optimized. The optimization procedure would have to take into account the practical concerns and exact model of the antenna and its orientation relative to the local ground.

\begin{figure}[tbh]
    \centering
    \includegraphics[width=0.49\linewidth]
    {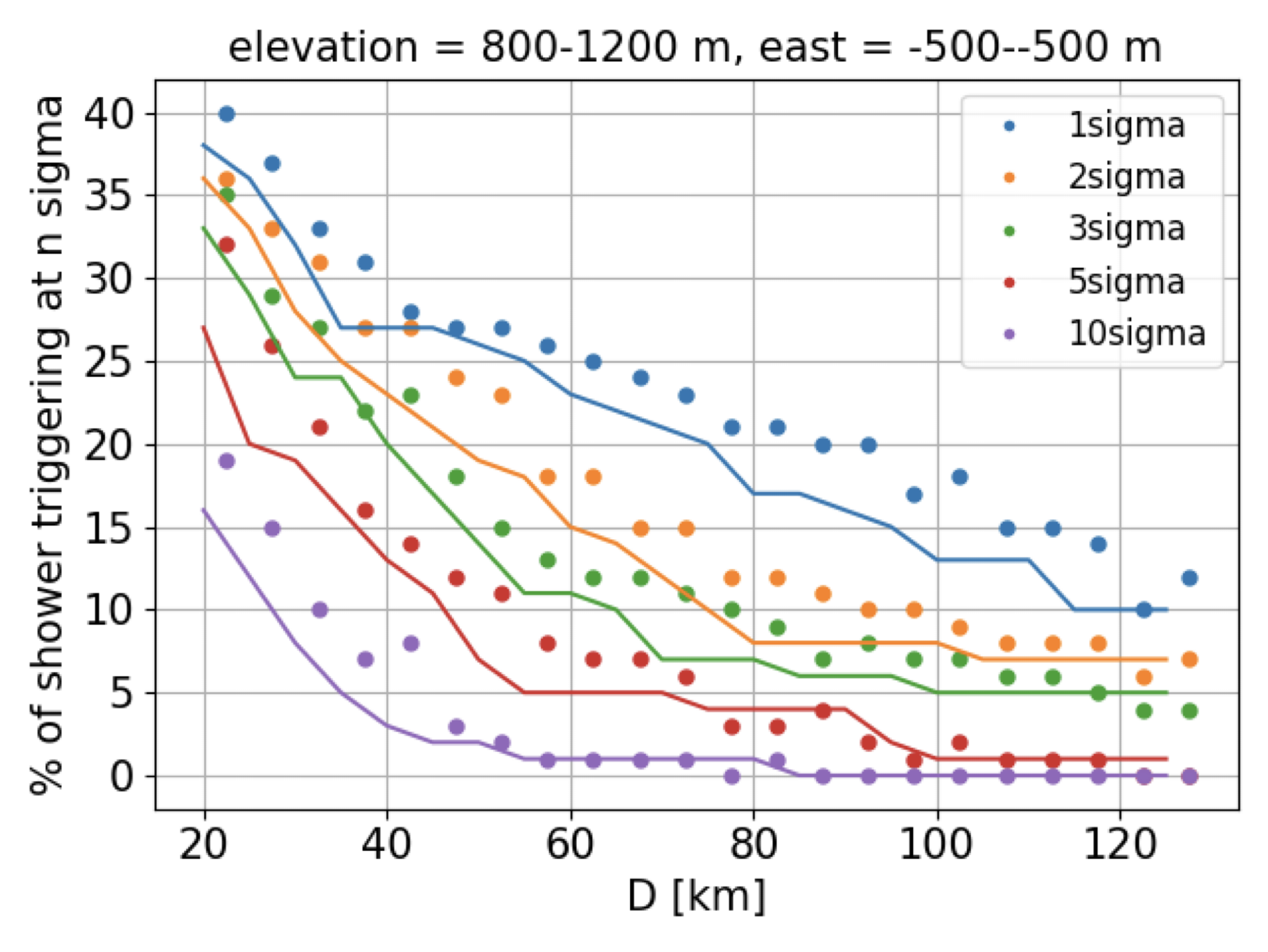}
        \includegraphics[width=0.49\linewidth]
    {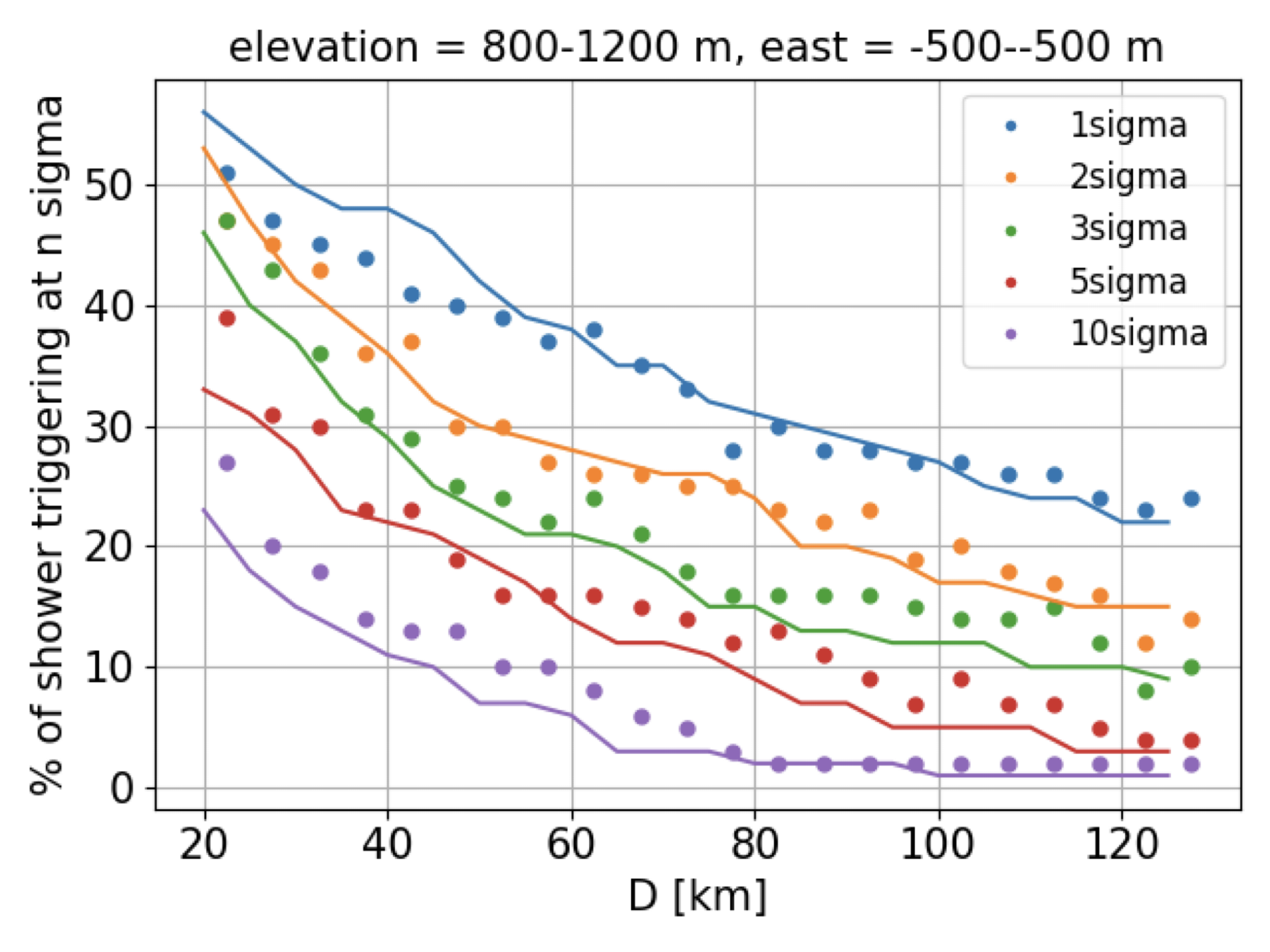}
    \caption{ Percentage of showers from an isotropic flux triggering at various thresholds above the irreducible noise backgrounds in the 50-200\,MHz band (22 $\mu V$) at varying distances, $D$, from the shower core and at location N\,$42.55^{\circ}$, E\,$86.88^{\circ}$. Electric fields modeled using ZHAiReS \cite{Alvarez_Mu_iz_2012} from air showers ($10^{17}$\,eV (left) and $10^{18}$\,eV (right)) modeled with both MARMoTs~\cite{Zeolla:ICRC, Zeolla:ARENA2024} (lines) and 
   a GRAND library~\cite{Decoene:2021atj} (dots).}
    \label{fig:simcomp}
\end{figure}

A conceptual design is shown in Fig.~\ref{fig:conceptdiagram}.  There are two sets of arrays:  triggering arrays surrounded by reconstruction arrays. Each triggering array is formed from compact phased arrays that lower the energy threshold, allow for tunable beams at the horizon, and directional radio frequency interference (RFI) rejection. The trigger arrays use low-gain and low frequency (30-80\,MHz) antennas that when phased together form higher gain beams. The reconstruction array is formed from a sparse array of autonomous antennas that provides long baselines needed for adequate reconstruction. They are also critical for RFI rejection as demonstrated with OVRO-LWA~\cite{Plant:ARENA2024} and TREND~\cite{Ardouin:2010gz}, which is particularly needed for arrays placed on elevated mountains. Because these sparse antennas cover a larger area, we also anticipate that they can enhance the sensitivity at higher energies. Using high-gain antennas tuned towards the horizon can enable reconstruction of the weak signals that trigger the phased arrays. Triggers are transmitted from the phased array to the reconstruction array.

We used simulation tools developed for both the GRAND~\cite{Decoene:2021atj} and BEACON experiments~\cite{Wissel_2020, Zeolla:ICRC, Zeolla:ARENA2024} to optimize the design. To get started we compared the electric fields and find them to be well within agreement. Showers generated by primary neutrinos of energies $10^{17}$\,eV and $10^{18}$\,eV are shown in Fig.~\ref{fig:simcomp} for the two simulation packages. Given the good agreement, we use the BEACON simulation package, MARMoTs~\cite{Zeolla:ICRC, Zeolla:ARENA2024}, to study the performance of an optimized design. In the present study, we assume a spherical Earth without topographical effects.

Based on the simulation studies presented in Secs.~\ref{sec:phased} and \ref{sec:sparse}, we assume a nominal design for the experiment. The triggering phased arrays use 24 dual-polarized dipoles based on the electrically short 30-80\,MHz dipoles used in the BEACON prototype~\cite{Southall:2022yil}. The mountain ridges are assumed to be aligned North to South, along the Earth's lines of longitude. The phased arrays are located at a prominence of 1\,km and are separated by 3\,km laterally along the mountain ridge. There are 10-20 autonomous antennas between each phased array, evenly spaced above and below the phased arrays. We assume that a trigger threshold of 5 times the noise given by Eqn.~\ref{eqn:ant_temp} on either the beams in the phased arrays or on the single autonomously triggered antennas in the reconstruction array. We also assume that the trigger is performed on the peak electric fields, which could either be achieved by aligning the array with the electric field or by a combined trigger on Vpol and Hpol.

\section{Low Thresholds with Compact Phased Arrays}\label{sec:phased}
\begin{figure}
    \centering
    \includegraphics[width=0.49\linewidth]{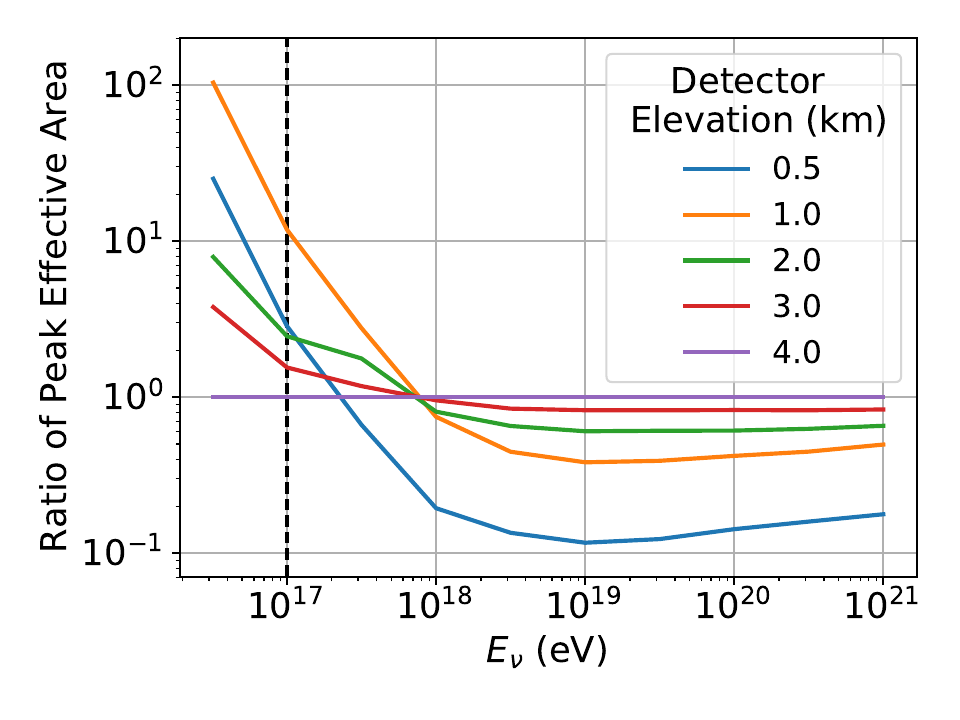}
    \includegraphics[width=0.49\linewidth]{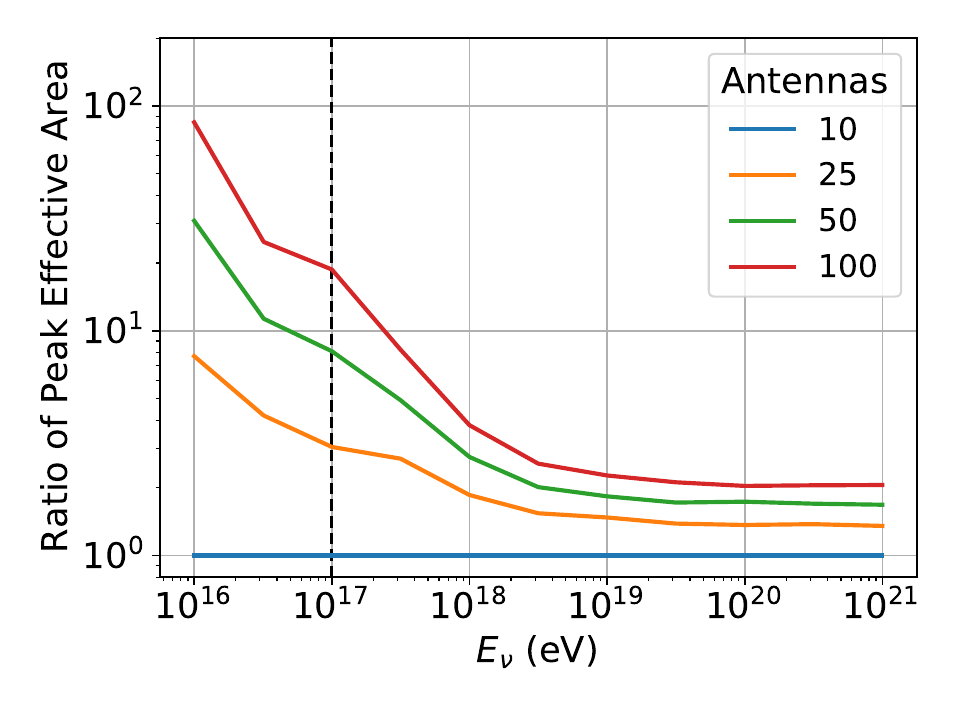}
    \caption{The impact on the peak effective area for varying detector prominence (left) and number of phased antennas included in the trigger (right). The ratio compares the peak effective area for each design to a prominence of 4.0\,km (left) and 10 phased antennas (right).}
    \label{fig:phasedarray_params}
\end{figure}
\begin{figure}
    \centering
    \includegraphics[width=0.8\linewidth]    {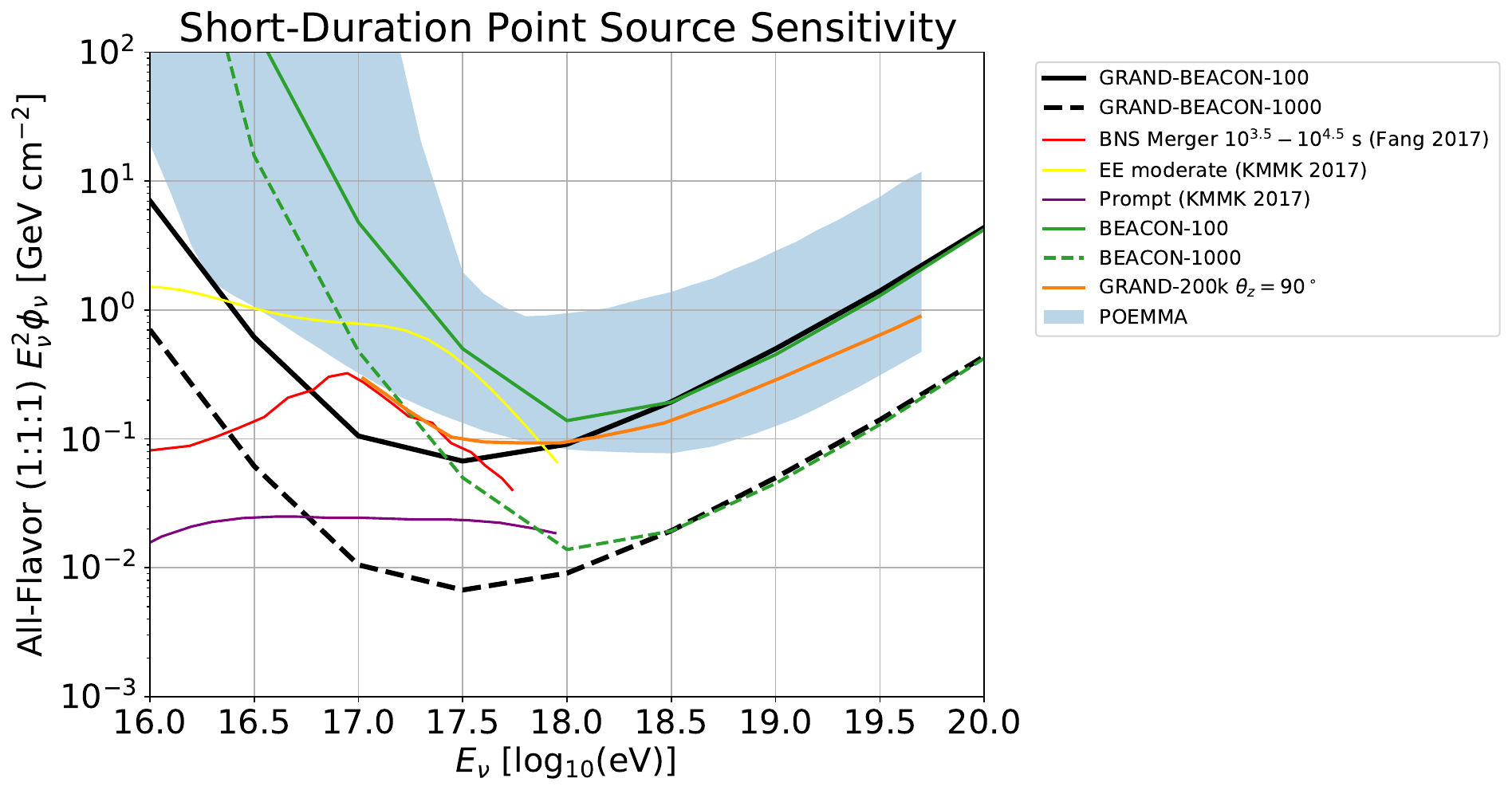}
    \caption{The instantaneous, transient sensitivity of the optimized GRAND-BEACON phased arrays (black curves) compared with both the BEACON (green), GRAND (orange), and POEMMA (blue band) concepts. The sensitivities are compared to the all-flavor fluence of neutrinos expected from a stable millisecond magnetar $10^{3.5} - 10^{4.5}$\,s after a BNS merger, scaled to a source distance of 5 Mpc \cite{Fang}, and the modeled all-flavor fluence from sGRB viewed on-axis, scaled to a distance of 40 Mpc, during two phases: extended (EE) and prompt emission \cite{KMMK}. }
    \label{fig:pointsource}
\end{figure}
We first considered the design for the triggering phased arrays, optimized for 100\,PeV. Fig.~\ref{fig:phasedarray_params} shows how the peak effective area changes when the detector elevation and number of antennas included in the phased array is adjusted.

The neutrino fluence sensitivity can be tuned by raising or lowering the detector elevation: higher detectors view a larger geometric area while lower detectors are sensitive to lower energy showers since the experiment is closer to the tau decay. If the elevation is too low (e.g. 0.5\,km), the detector is in the near field of the shower for most events which can reduce the sensitivity~\cite{Wissel_2020}. A detector prominence of 1\,km enhances the peak effective area by a factor of 10 compared with 4\,km at 100 PeV, while reducing the effective area at the highest energies by a factor of 2.

Increasing the number of antennas included in the phased array increases the peak effective area. 
We chose 24 antennas for our fiducial design because it enhances the effective area at 100 PeV by a factor of 3 and represents a tractable number of antennas to include in a small area ($<$100-200\,m between the longest two baselines). It is also feasible to include in a present-day FPGA that could be used to form the phased array trigger and digitize the waveforms. We note that we assume perfect phasing for these simulations, meaning that the antennas used in the triggering array must be placed close enough together that they view the same portion of the shower. 

An example of the power of this optimized approach is shown in Figs.~\ref{fig:pointsource} for 100 and 1,000 phased array stations, respectively. In this model, we assume an optimistic design with 1\,km detector prominence, 24 antennas per phased array, 6 dBi gain which is perfectly matched across a band of 30-80\,MHz, a sky fraction of 30\%, and $\mathcal{F}=0\%$ ground reflection. The stations are centered on the location of the BEACON prototype $(\lambda=37.5893^\circ \, N, \, \phi=118.2376^\circ \, W)$. In this scenario, the instantaneous point source sensitivity improves over the GRAND and BEACON designs by an order of magnitude at 100\, PeV. This can improve further when the reconstruction antennas also contribute to the trigger (as discussed in Sec.~\ref{sec:sparse}). Given that the equatorial latitudes assumed for a tau neutrino detector such as this has broad sky coverage when integrated over a day, this instrument is optimized for short-duration transients triggered by current and upcoming time-domain electromagnetic experiments. 
Triggered multimessenger searches, particularly if they are stacked, can lead to the first discoveries of UHE neutrinos.

\section{Reconstruction with Sparse Antennas}\label{sec:sparse}

The reconstruction array would include 10-20 antennas spaced between the denser triggering arrays. The exact placement and number of antennas needed is yet to be determined, but will depend on how they impact the reconstruction of the radio arrival direction, the discrimination between cosmic rays and neutrinos, and the effective area at high energies. The angular resolution achieved on the radio arrival direction will impact the reconstruction of the neutrino arrival direction and the rejection of any anthropogenic radio backgrounds in the field of view of the experiment.

The triggering arrays are nominally separated by 3\,km -- although this is another parameter that could be optimized --- and the reconstruction antennas are placed above and below the phased arrays on the mountain slope to further improve zenith angle reconstruction. The reconstruction antennas will trigger autonomously and receive a lower-threshold trigger from the phased arrays. The phased arrays will send an external trigger via long-range WiFi. The clocks on both the autonomous antennas and the phased array data acquisition system can be synchronized by phase-aligning a transmitted sine wave as demonstrated previously by AERA~\cite{PierreAuger:2015aqe}. 

\begin{figure}
    \centering
    \includegraphics[width=0.9\linewidth]{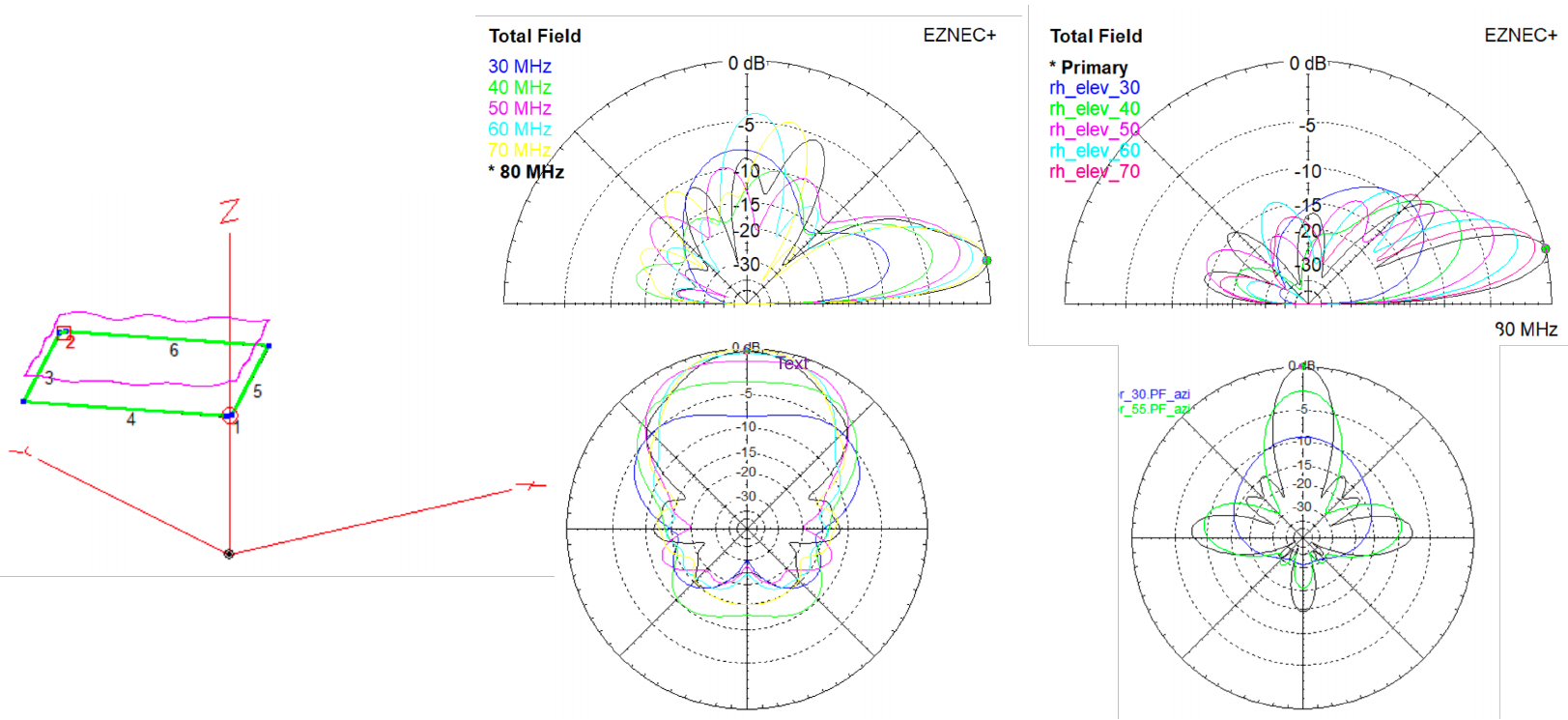}
    \caption{Design of the rhombic antenna under consideration for the high gain antennas in the reconstruction array, shown in the Hpol configuration (left). Realized gain, both in the direction from the local zenith to the horizon ($\theta$, top) and parallel to the local ground ($\phi$, bottom), are shown for Vpol (middle) and Hpol (right).}
    \label{fig:rhombic_ant}
\end{figure}

Because the primary purpose of the reconstruction antennas is to enable neutrino identification and reconstruction for the lower threshold signals triggered by the phased array, the weak signal from the autonomous antennas needs to be extracted from noise. We anticipate using interferometric reconstruction~\cite{Schoorlemmer:2020low,Schluter:2021egm} combining the signals from both the reconstruction array and the phased array to reconstruct the radio arrival direction across the km-scale baselines. Similarly, the autonomous antennas should also have a higher gain design than the individual antennas in the phased array. One possible antenna design is shown in Fig.~\ref{fig:rhombic_ant}.  A rhombic antenna consists of a long wire arranged in a rhombus that can be readily pointed towards the horizon. For Vpol, only one 5-m tall mast and a 10-m long wire is needed, assuming that the mountain slopes down. For Hpol, four 4-m tall masts would be needed. In Vpol, the peak realized gain is 4.5\,dBi which can be readily pointed towards or below the horizon with a 15$^{\circ}$ slope. In Hpol, the peak realized gain is 14\,dBi. Other options for high gain antennas include LPDAs and stacked SALLA antennas, but the former are challenging to mount in high winds at low frequencies and the later have substantial resistive losses. 

\begin{figure}
    \centering
    \includegraphics[width=0.49\linewidth]{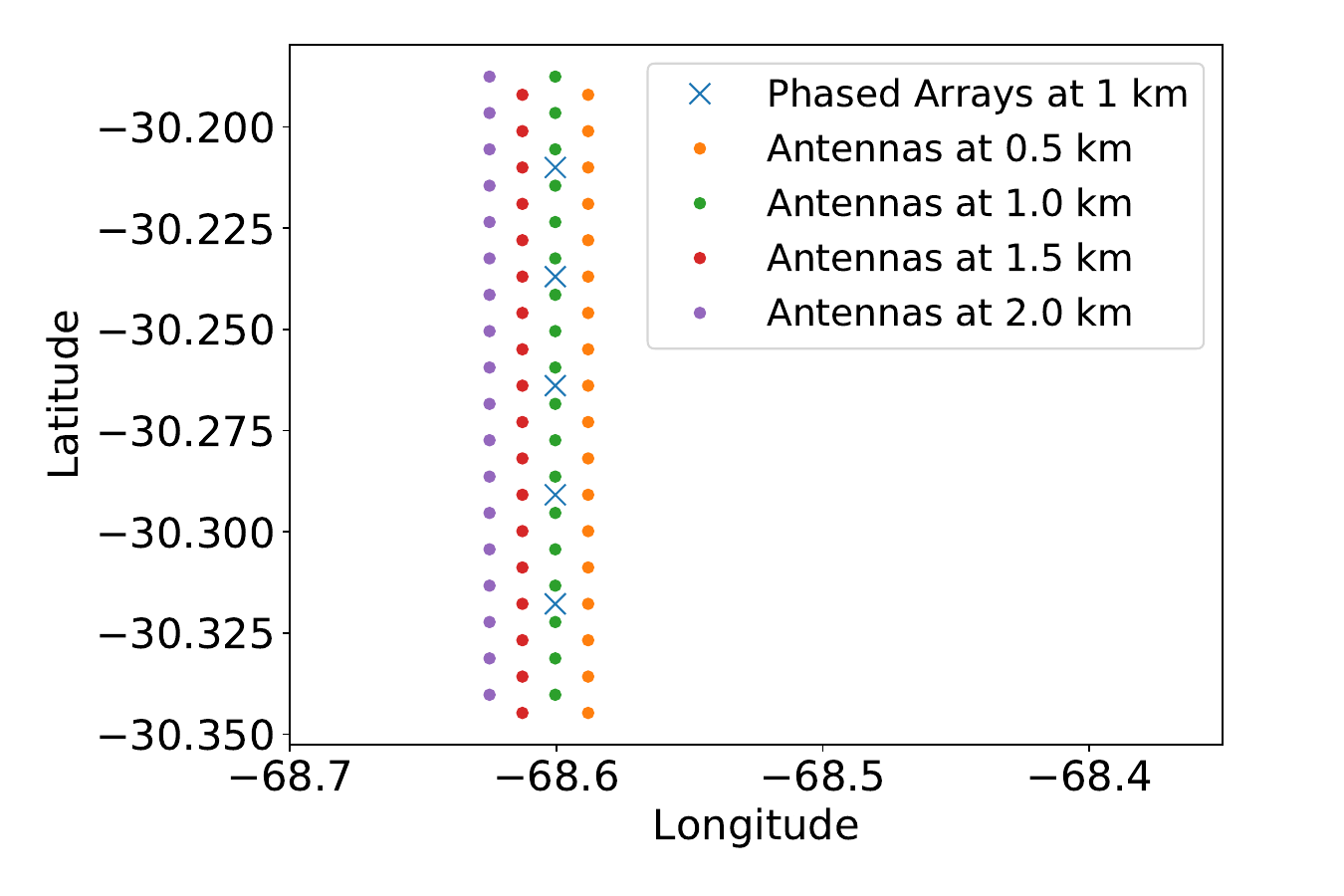}
    \includegraphics[width=0.44 \textwidth]{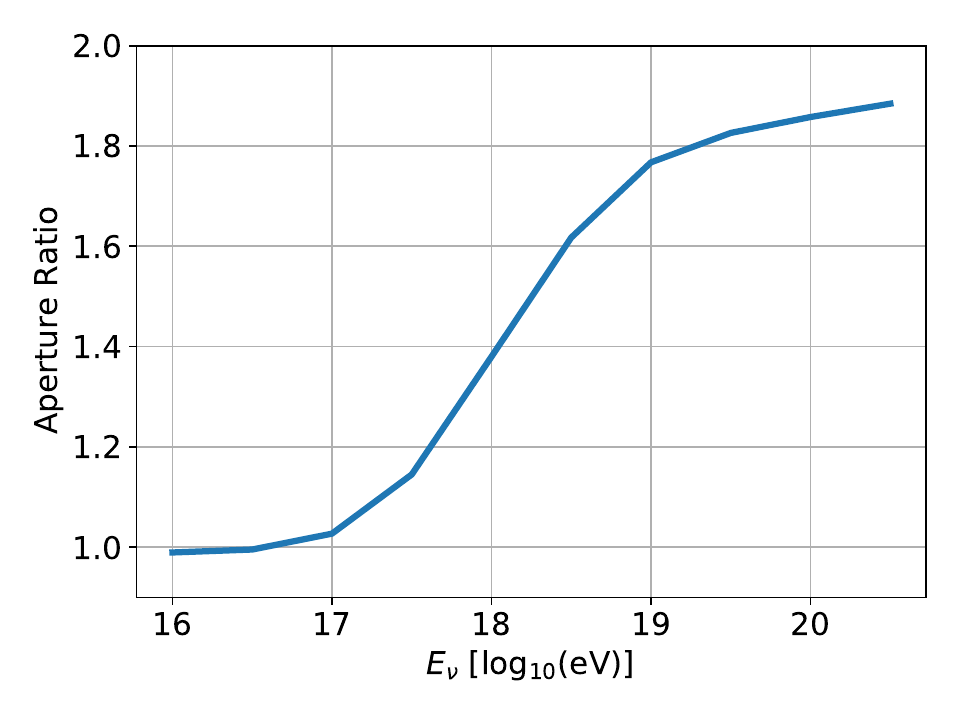}
    \caption{Five 24-channel phased arrays and 72 autonomous antennas in San Juan, Argentina (left). Aperture ratio using both the phased arrays and the autonomous antennas compared to only the phased arrays (right).}
    \label{fig:sparse_effarea}
\end{figure}

The autonomously triggering antennas in the reconstruction array naturally add to the sensitivity of the experiment by extending the range of the events visible at higher energies. We simulated an array of five 24-channel phased arrays with 14 higher-gain antennas\footnote{Modeling higher gains using the Vpol rhombic antenna gain applied to the full electric fields.} 
between each pair of phased arrays. We then compared the aperture (i.e. the effective area integrated over solid angle) for this design to the phased arrays alone. The resulting increase in the aperture is shown in Fig.~\ref{fig:sparse_effarea}. By including less than a factor of two more autonomous antennas, the aperture grows by nearly a factor of two at the highest energies. This balances the loss of high energy sensitivity by placing the detectors at a lower elevation than in the BEACON design, discussed in Sec.~\ref{sec:phased}.

\section{Conclusions}

The combined approach of phased triggering arrays and autonomous and sparse reconstruction antennas has several advantages. The phased arrays allow the experiment to target lower energies than in either the BEACON or GRAND designs. The beams can be trained directly on the horizon and directional masking of individual beams can be used for background rejection. The sparse antennas are spread out across longer baselines. As such, they can enhance event reconstruction, discrimination between neutrinos and cosmic rays and other backgrounds. They also provide additional effective area at high energies. Future work will focus on further optimizing the design, including the local topography, and quantifying the reconstruction capability of the array. 

\setlength{\bibsep}{0pt plus 0.3ex}
\bibliographystyle{JHEP}
{\footnotesize
\bibliography{bib}
}

\section*{Full Author List}
\noindent Stephanie Wissel\textsuperscript{1,3,4,5}
Andrew Zeolla\textsuperscript{1,3,4}
Cosmin Deaconu\textsuperscript{1,6}
Valentin Decoene\textsuperscript{1,2,7}
Kaeli Hughes\textsuperscript{1,8}
Zachary Martin\textsuperscript{1,6}
Katharine Mulrey\textsuperscript{1,9}
Austin Cummings\textsuperscript{1,3,4,5}
Rafael Alves Batista\textsuperscript{2,10,11}
Aurélien Benoit-Lévy\textsuperscript{2,12,3,4}
Mauricio Bustamante\textsuperscript{2,13}
Pablo Correa\textsuperscript{2,11}
Arsène Ferrière\textsuperscript{2,12}
Marion Guelfand\textsuperscript{2,10,11}
Tim Huege\textsuperscript{2,14,15}
Kumiko Kotera\textsuperscript{2,3,4,10,16}
Olivier Martineau\textsuperscript{2,10,11}
Kohta Murase\textsuperscript{2,3,4,5,6}
Valentin Niess\textsuperscript{2,17}
Jianli Zhang\textsuperscript{2,18}
Oliver Krömer\textsuperscript{19}
Kathryn Plant\textsuperscript{20}
Frank G. Schroeder\textsuperscript{21,14}
\\
\\
\noindent\textsuperscript{1} for the BEACON Collaboration\\
\textsuperscript{2} for the GRAND Collaboration\\
\textsuperscript{3}Center for Multi-Messenger Astrophysics, Institute for Gravitation and the Cosmos, Pennsylvania State University, University Park, PA 16802\\
\textsuperscript{4} Dept. of Physics, Pennsylvania State University, University Park, PA 16802\\
\textsuperscript{5} Dept. of Astronomy and Astrophysics, Pennsylvania State University, University Park, PA 16802\\
\textsuperscript{6} Dept. of Physics, Enrico Fermi Institute, Kavli Institute for Cosmological Physics, University of Chicago, Chicago, IL 60637\\
\textsuperscript{7}SUBATECH, IN2P3-CNRS, UMR 6457, Nantes Université, Ecole des Mines de Nantes, 4 rue Alfred Kaslter, Nantes, France\\
\textsuperscript{8}Department of Physics, Center for Cosmology and AstroParticle Physics, The Ohio State University, OH, USA, 43210\\
\textsuperscript{9}Department of Astrophysics/IMAPP, Radboud University, P.O. Box 9010, 6500 GL Nijmegen, The Netherlands\\
\textsuperscript{10}Sorbonne Université, CNRS, UMR 7095, Institut d'Astrophysique de Paris, 98 bis bd Arago, 75014 Paris, France\\
\textsuperscript{11}Laboratoire de Physique Nucléaire et des Hautes Energies (LPNHE), 4 Pl. Jussieu, 75005 Paris, France\\
\textsuperscript{12}Université Paris-Saclay, CEA, List, F-91120, Palaiseau, France\\
\textsuperscript{13}Niels Bohr International Academy, Niels Bohr Institute, University of Copenhagen, DK-2100 Copenhagen, Denmark\\
\textsuperscript{14}Institute for Astroparticle Physics (IAP), Karlsruhe Institute of Technology, 76021 Karlsruhe, Germany\\
\textsuperscript{15}Astrophysical Institute, Vrije Universiteit Brussel, 1050 Brussels, Belgium\\
\textsuperscript{16}Physics Department, Vrije Universiteit Brussel, Pleinlaan 2, 1050 Brussels, Belgium\\
\textsuperscript{16}Center for Gravitational Physics, Yukawa Institute for Theoretical Physics, Kyoto, Kyoto 606- 8502 Japan\\
\textsuperscript{17}CNRS/IN2P3 LPC, Université Clermont Auvergne, F-63000 Clermont-Ferrand, France\\
\textsuperscript{18}National Astronomical Observatories, Chinese Academy of Sciences, Beijing 100101, China\\
\textsuperscript{19}Institute for Data Processing and Electronics (IPE), Karlsruhe Institute of Technology, 76021 Karlsruhe, Germany\\
\textsuperscript{20}Jet Propulsion Laboratory, California Institute of Technology, 4800 Oak Grove Drive, Pasadena, CA, USA\\
\textsuperscript{21}Bartol Research Institute, Department of Physics \& Astronomy, University of Delaware, Newark, DE, USA
\end{document}